\documentclass[prl,twocolumn,nofootinbib,aps,superscriptaddress,showpacs]{revtex4-1}
\usepackage{color,graphicx,shortvrb,epsfig}

\usepackage{graphicx}
\usepackage{dcolumn}
\usepackage{bm}

\usepackage{colordvi}
\usepackage{psfig,epsfig,amssymb,amsmath,latexsym}

\newcommand{\no}{\nonumber}
\def\lsim{\mathrel{\rlap{\lower4pt\hbox{\hskip1pt$\sim$}}
    \raise1pt\hbox{$<$}}}         
\def\gsim{\mathrel{\rlap{\lower4pt\hbox{\hskip1pt$\sim$}}
    \raise1pt\hbox{$>$}}}         

\def\lsim{\mathrel{\rlap{\lower4pt\hbox{\hskip1pt$\sim$}}
    \raise1pt\hbox{$<$}}}         
\def\gsim{\mathrel{\rlap{\lower4pt\hbox{\hskip1pt$\sim$}}
    \raise1pt\hbox{$>$}}}         

\def\beq{\begin{equation}}
\def\eeq{\end{equation}}
\def\ba{\begin{eqnarray}}
\def\ea{\end{eqnarray}}

\def\<{\langle}
\def\>{\rangle}

\begin{document}

\begin{flushright}
\date{}
\end{flushright}

\title{Statistical
mechanics of bent twisted ribbons}
\author{Supurna Sinha}
\author{Joseph Samuel} 
\affiliation{Raman Research Institute, Bangalore, India 560 080}


\begin{abstract}
We present an analytical study of bent twisted ribbons. 
We first describe the elastic response of a ribbon 
within a purely mechanical framework. We then
study the role of thermal fluctuations in 
modifying its elastic response. 
We predict the moment angle relation of bent and twisted ribbons. 
Such a study is expected to shed light on the  role of twist in 
the ``J factor'' for 
DNA looping and on bending elasticity of twisted graphene
ribbons.  

\end{abstract}

\maketitle


DNA is a twist storing polymer. Many biological processess
in the cell exert torsional stresses on 
the DNA molecule.
One would expect that such torsional stresses affect the bending
elasticity of the molecule and its ability to form loops. ``Cyclization''
of DNA has been the focus of several studies, theoretical and 
experimental, {\it in vitro} and {\it in 
vivo}\cite{nelsonloop,eplac,shimada,cloutier}.
A proper understanding of DNA elasticity over a range of length scales
is valuable to understanding biological processes. 
The persistence length $L_P$ of DNA 
is about $50 nm$. At length scales short compared to $L_P$ the molecule
is well described by a purely energetic treatment similar to that
used by civil engineers to describe the twisting and bending of 
beams and cables\cite{landau,love,class}. 
At length scales of the order of the persistence
length, the effect of thermal fluctuations becomes appreciable. 
The subject of this study is the role of thermal 
fluctuations in shaping the elastic properties of torsionally stressed 
macromolecular beams. Our study is general and applies equally 
to other systems like actin filaments or carbon nanotubes 
under mechanical stresses\cite{carbon}.
We use an approximation scheme 
\cite{aghosh,polley,fluct} developed earlier
to account for thermal fluctuations using the Van Vleck correction.
Such an approximation works well in the stiff limit and 
as described earlier, works\cite{polley} suprisingly well even 
over length scales as large as five times the persistence length.
We work within the wormlike chain model\cite{kratkyporod,doi},
which views the molecule as a ribbon, carrying an energy cost for
bending and twisting. 
The wormlike chain  model has been known to describe
double stranded DNA \cite{siggia} as well as actin 
filaments\cite{wilhelmfrey}. Fig. 1 shows typical 
configurations of a twisted bent ribbon.
\begin{figure}[h!t]
\includegraphics[width=8.0cm]{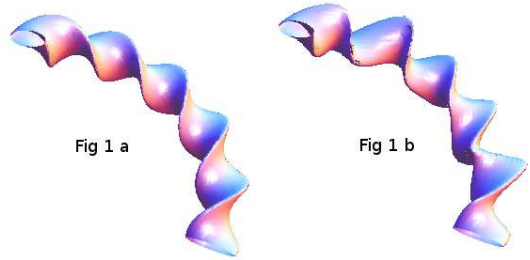}
\caption{Figure shows two configurations of a twisted and bent molecule 
with  
fixed end tangent vectors $\hat{t}_i$ and $\hat{t}_f$. 
On the left (Fig. 1a) is the minimum energy configuration and on the right
(Fig. 1b) the same configuration is shown slightly perturbed by thermal fluctuations.
}
\label{cartoon}
\end{figure}

For clarity we consider experiments
in which one end of the molecule is tethered to a glass slide, its tangent
vector at the same end is constrained to lie 
along the $\hat{t}_i$ direction and the molecule is torsionally
constrained at both ends. 
We can attach a magnetic bead to the untethered end and
apply bending moments and twist up the molecule
by varying the direction of an applied magnetic field. 
Note that we do not constrain the final position of the molecule $x(L)$ 
but only its final tangent
vector $\hat{t}_f$.
A uniform magnetic field will result in a pure bending moment 
without applying any stretching 
force. Plotting the bending moment vs the 
bending angle gives another experimental
probe of the elastic properties. In this paper we derive the predictions of
the wormlike chain for such experimental situations.

We present an approximate analytical 
solution of this statistical mechanical problem.
Working with the wormlike chain model, 
we first write down a closed form  analytical solution in parametric form,
expressing the mechanical energy functional of a 
twisted bent ribbon as a function of the bending angle. We use this to
plot the theoretically predicted moment-angle relation in the absence
of thermal fluctuations. We then compute the
fluctuation determinant which describes the leading thermal correction
to the energy functional. The fluctuation determinant is used to discuss
the stability of the classical solutions.
Finally, we describe how thermal fluctuations modify the moment angle 
relations and conclude with a
discussion.

In the wormlike chain,
we model the polymer by a ribbon, a 
framed space curve $\{\vec{x}(s),\hat{e}^i(s)\}$. 
 $\vec{x}(s)$ describes 
the  curve, $\hat{t}(s)  =  \frac{d\vec{x}}{ds}$, 
its tangent vector  and ${\hat e}^{i}(s)$ the framing (the ${\hat e}^i$ s are an
orthonormal frame with ${\hat e}^3={\hat t}$). 
$s$ is the arc length parameter along the curve ranging from $0$ to $L$, the 
contour length of the curve. 
${\vec x}(0)=0$ since one end 
is fixed at the origin. The tangent 
vectors at both ends ${\hat t}(0)$,${\hat t}(L)$ are fixed to $\hat{t}_i$ and
$\hat{t}_f$ respectively. 

By simple transformations described elsewhere 
(\cite{fluct,samsupabhi,papa})
we can reduce this problem by eliminating the twist degree of freedom.
The reduced problem deals with a space curve ${\vec x}(s)$ with tangent
vector ${\hat t}(s)$ subject to a writhe constraint.
The  energy functional is 
\begin{equation}
{\cal E} ({\cal C}) = \frac{A}{2} \int^{L}_{0} 
(\frac{d\hat{t}}{ds}.\frac{d\hat{t}}{ds}) ds-2\pi \tau {\cal W},
\label{energy}
\end{equation}
where $A$ is an  elastic constant with dimensions 
of energy times length. The quantity ${\cal W}$ is the writhe
which is constrained to take a fixed value.  The constraint is enforced 
by the Lagrange multiplier $\tau$, which
has the physical interpretation of the torque imposed on the ribbon.
(The writhe of an open space curve whose initial and final tangent
vectors are fixed is defined \cite{papa} by extending the curve 
beyond its ends to infinity adding straight
line segments with constant tangent vectors ${\hat t}_i$ and ${\hat 
t}_f$.)
We will work in the constant torque ensemble, where $\tau$ is held
fixed.
The quantity $L_p=A/kT$ is the persistence length. 
The mathematical problem we face is to compute the partition function
\begin{equation}
Q(\hat{t}_i,\hat{t}_f)= \sum_{{\cal C}} \exp - \Big[\frac{{\cal E}({\cal 
C})}{k_B T}\Big]\;. 
\label{ptnfull}
\end{equation}
In Eq.(\ref{ptnfull}), the sum is over all allowed configurations of the 
polymer, those which satisfy the boundary conditions
for the tangent vector at the two ends: 
$\hat{t}_i=\hat{t}(0),\hat{t}_f=\hat{t}(L)$.

In the stiff limit, one can neglect thermal fluctuations and 
simply minimize the energy functional appearing in (\ref{ptnfull}).
The problem is formally similar to that of a symmetric top, 
(as was known to Kirchoff\cite{kirchoff}) and this greatly aids the 
solution.
Borrowing from classical mechanics, we can use variational techniques and
identify ``constants of the motion''
The Euler-Lagrange (E-L) equations that one arrives at from the 
variational problem are\cite{papa}:
\begin{eqnarray}
\ddot{\hat t} =-\tau({\hat t}\times{\dot{\hat t}})+\gamma{\hat t},
\label{eqnofmot} 
\end{eqnarray}
where the term $\gamma{\hat t}$ arises since $\delta{\hat t}\cdot{\hat
t} = 0$. 
The total ``energy'' 
\begin{eqnarray}
{\cal H}=\frac{1}{2} 
(\dot{\theta}^2+\sin^{2}\theta\dot{\phi}^2)
\label{energyex}
\end{eqnarray}
is a ``constant of the motion'' as is the ``angular
momentum''
\begin{eqnarray}
\vec{J}=({\hat t} \times\dot{\hat t}\,)-\tau{\hat t} .
\label{ang}
\end{eqnarray}
For positive (negative) $\tau$,
we choose the 
$-{\hat z}$  (${\hat z}$) axis along the 
direction of the angular momentum $\vec{J}$. Below we restrict to
$\tau$ positive. Similar considerations apply to negative $\tau$ with
appropriate changes.
We have 
\begin{equation} 
-J=J_z=\sin^{2}\theta\dot{\phi}   - \tau\cos\theta,  
\label{jz}
\end{equation}
where we have introduced the usual polar coordinates on the space of
tangent vectors 
${\hat t}=(\sin{\theta}\cos{\phi},\sin{\theta}\sin{\phi},\cos{\theta})$.
Evaluating $-\hat{t}.\vec{J}=\tau=J\cos{\theta}$ we find
that the angle $\theta$ is a constant of the motion and that $\dot{\phi}=-J$.
$\mu=\cos{\theta}$ is positive (for $\tau>0$, a condition we assume).
 The solution of the Euler-Lagrange equation is ${\hat t}_{cl}(s)=
(\sin{\theta} \cos{J s},-\sin{\theta} \sin{J s},\cos{\theta})$.
The tangent vector is restricted to a fixed latitude on the 
unit sphere (corresponding to ``precession without nutation'' 
in the top analogy). In real space 
the molecule describes a helix.

By choice of the $x$ axis we set the initial and final tangent vectors to:
${\hat t_i}=(\sin{\theta},0,\cos{\theta})$ and
${\hat 
t_f}=(\sin{\theta}\cos{\phi_f},\sin{\theta}\sin{\phi_f},\cos{\theta})$.
Considering the dot product
$u = {\hat t_i} . {\hat{t}_f} = \cos{\alpha}$ and setting $\mu = \cos{\theta}$
gives us $u$ and therefore $\alpha$ as a function of $\mu$. $u(\mu)$
is plotted in Fig. 2.
\begin{eqnarray}
u(\mu)=(1-\mu^2)\cos(\frac{\tau L}{\mu})+\mu^2 
\label{uexpress}
\end{eqnarray}
and an expression for $\alpha$ as a function of $\mu$.
\begin{eqnarray}
\alpha(\mu)=\arccos{\big[(1-\mu^2)\cos(\frac{\tau L}{\mu})+\mu^2}\big] 
\label{alphaexpress}
\end{eqnarray}

For fixed $\alpha$, $u=\cos{\alpha}$ defines a horizontal straight line
(shown in black in Fig. $2$), which intersects the curve (Eq. 
(\ref{uexpress})) at multiple vallues of $\mu$. The largest of these
corresponds to the lowest value of $J$. The smaller values of $\mu$ 
correspond to higher $J$ values and higher rates of traversal, including
traversing the circle of latitude more than once. As we will see below,
only the largest $\mu$ value gives a stable configuration. 

\begin{figure}[h!t]
\includegraphics[width=8.0cm]{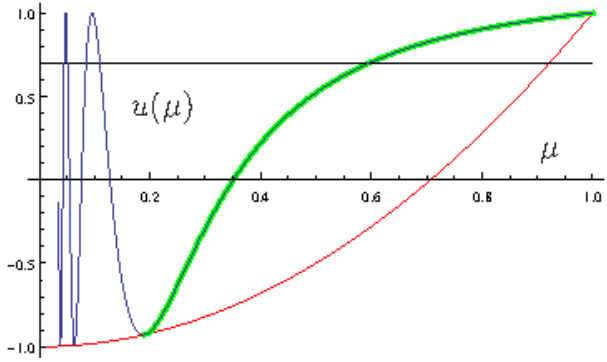}
\caption{A plot of $u(\mu)$ versus $\mu$. The black horizontal line is
a line of fixed $\alpha$. The red parabolic line represents $2\mu^2-1$
which is the envelope of the curve $u(\alpha)$ near its minima. Note
that the minima are very close to the points of tangency between the
envelope and $u(\alpha)$.}
\label{uofmu}
\end{figure}

Note that ${\hat t}_{cl}$ remains in the northern hemisphere since $\mu>0$.
Such curves (and their small perturbations) are ``good curves'' 
\cite{papa}, in the sense that they can 
be continuously deformed to the straight line along ${\hat z}$ without
self intersection or pointing towards $-{\hat z}$. We can therefore use
Fuller's formula for the writhe $2\pi{\cal W}=\int_0^Ld\phi(1-\cos{\theta})$
and compute the energy functional in terms of $\mu$
by doing the integrations in (\ref{energy}) for the classical solution:
\begin{eqnarray}
{\cal E}_{cl}(\mu)=\frac{\tau^2 L}{2}[(\frac{1}{\mu} + 1)^2-4]
\label{energyexpress}
\end{eqnarray}
Equations ($\ref{alphaexpress}$) and ($\ref{energyexpress}$)
give the solution ${\cal E}_{cl}(\alpha)$
in parametric form. Differentiating ${\cal E}_{cl}(\alpha)$ with respect to 
$\alpha$ gives us the  moment-angle relation:
\begin{equation}
M_{cl}(\alpha)=\frac{\partial {\cal E}_{cl}}{\partial \alpha}
\label{momang}
\end{equation}
where the subscript ${cl}$ on the left hand side of (\ref{momang})
reminds us that this expression ignores thermal flucuations.

So far, we have dealt with the classical solutions of the Euler-Lagrange
equations, ignoring questions of stability and thermal fluctuations. Both
of these questions will now be addressed. As explained in \cite{fluct} in
great detail, the fluctuation determinant $\Delta$ can be calculated from the 
Energy functional ${\cal E}_{cl}({\hat t}_i,{\hat t}_f)$ 
using the Van Vleck determinant
\begin{equation}
\Delta^{-1}= det \frac{\partial^2 {\cal E}_{cl}}{\partial {\hat t}_i \partial {\hat t}_f}
\label{det}
\end{equation}
This computation of this two by two determinant yields an elegant formula
for the inverse of the fluctuation determinant:
\begin{equation}
\Delta^{-1}=\frac{1}{\sin{\alpha}} \frac{\partial}
{\partial \alpha} \big[M_{cl}^2(\alpha)\big]
\label{elegant}
\end{equation}
where $M_{cl}(\alpha)$ the classical Moment-angle relation. 
Eq. (\ref{elegant}) is
the main result of this paper can be used to understand 
both the stability and the fluctuations around
the classical solutions. Eq. (\ref{elegant}) can be used to graphically 
display
the free energy and moment angle relation of the ribbon. 

The fluctuation determinant is the product of the eigenvalues of
the fluctuation operator. A stable configuration is a local minimum
of the energy and so the fluctuation operator has only positive eigenvalues.
Instability sets in when at least one of the eigenvalues crosses zero. At
such points the fluctuation determinant $\Delta$ vanishes. Thus, $\Delta$
Eq. (\ref{elegant}) can be used to study the stability of classical solutions.
Note first that $\mu=1$ describe a straight line configuration. These
configurations are known to be stable  for
$\tau< \pi/L$ \cite{love,papa}, a condition we will assume hereafter. 
Suppose
$0<\mu<1$, so that $\alpha\neq 0,\pi$. At points $\mu_0$ 
where $du/{d\mu}(\mu_0)\neq 0$, Taylor expansion reveals that
$u(\mu),{\cal E}_(\mu)$ and $\alpha(\mu)$ vary linearly in $(\mu-\mu_0)$.
A small calculation shows that $\Delta^{-1}$ is finite and 
so $\Delta$ does not vanish. It follows that configurations remain
stable as long as $du/{d\mu}(\mu_0)\neq 0$. Instability sets in
when
\begin{equation}
\frac{d u(\mu)}{d \mu}=0
\label{instability}
\end{equation}
From Fig. 2 we see that the green region is stable but the remaining
configurations are not. Eq. (\ref{instability}) is a transcendental equation
which can be solved numerically. A glance at Fig. 2 reveals that the 
roots of this equation are close to (but not equal to) the points where
$u(\mu)$ touches the envelope (shown in red online) $2\mu^2-1$, where 
$\alpha=2\theta$. For a ribbon bent by angle $\alpha$ and subject to a torque
$\tau$, the critical line for buckling is close to  $\tau=\pi/L \cos{\alpha/2}$, 
the exact value being given by the roots of Eq. (\ref{instability}). 
For torques and bending angles above the critical line the molecule
buckles into a loop and cyclizes. It is evident from this formula that
higher torque implies lower $\alpha$ for buckling. Thus torque aids
cyclization as could be naively expected.

Another factor that aids cyclization is thermal fluctuations.
The thermal fluctuations are easily incorporated in 
using Eq. (\ref{elegant}) in the expression for the free energy
\begin{equation}
{\cal F}(\alpha(\mu)) ={\cal E}_{cl}(\mu)+1/2 kT \log{\Delta(\mu)}
\label{helmholtz}
\end{equation}
This free energy can be differentiated to give the moment angle
relations when thermal fluctuations are incorporated. 
The results are best
seen in graphical form. Fig. 3  shows the Helmholtz energy as a function
of $\alpha$ and Fig.4 shows the Moment-Angle relations with and without
thermal fluctuations.

\begin{figure}[h!t]
\includegraphics[width=8.0cm]{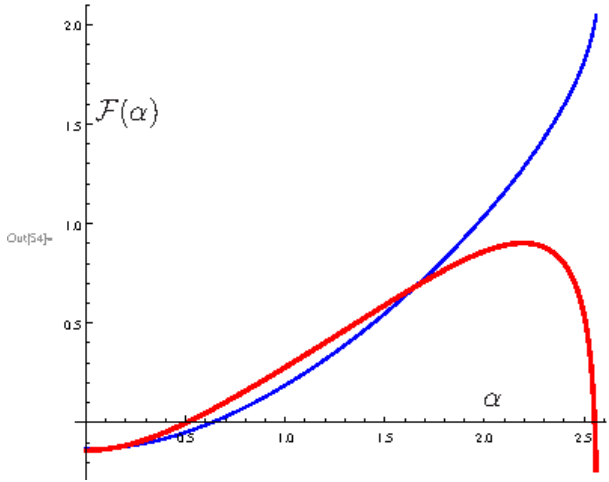}
\caption{ The Helmholtz free energy ${\cal F}(\alpha)$
versus $\alpha$ for parameters $\tau=.3,L=3,kT=.5$, 
with (thick red curve) and without thermal fluctuations (thin blue curve).}
\label{helmholtzfig3}
\end{figure}

\begin{figure}[h!t]
\includegraphics[width=8.0cm]{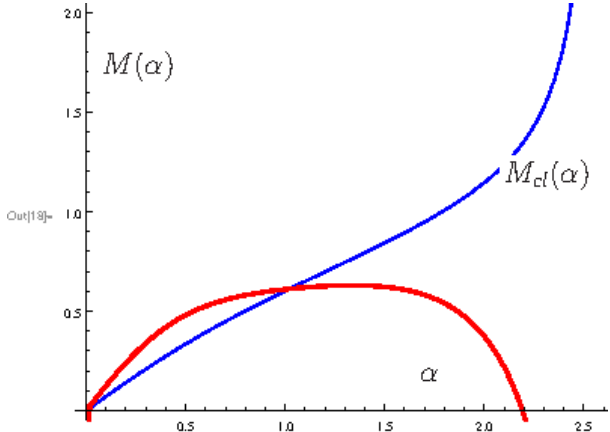}
\caption{Bending moment versus bending angle for the same parameter values
as in Fig.3 caption ($\tau=.3,L=3,kT=.5$).
Thin blue line 
shows the mechanical elastic response
and the thick red line includes the effect of thermal fluctuations.
}
\label{momentangle}
\end{figure}

To conclude, in this paper we have studied the elastic response of a 
bent twisted ribbon and have analyzed how thermal fluctuations modify
its response. Note that the $\tau \rightarrow 0$ limit is a singular one
and has to be taken with care. From $\tau = J \mu$ it follows that 
$\tau \rightarrow 0$ also implies $\mu \rightarrow 0$. Both 
$\tau$ and $\mu$ need to be taken to $0$ in such a way as to keep
$J$ finite. We find that classically the energy is quadratic for 
small values of the bending angle and then for large bending angles 
it diverges. Correspondingly, the moment is initially Hookean and
then diverges for large values of the bending angle.

In the presence of thermal fluctuations there is a softening of the 
``bending rigidity'' which is reflected in the curves depicting 
the free energy and the moment angle relations. This has implications
for cyclization and DNA looping ``J factors''. Our analysis suggests 
that in a cellular environment,
torsional stresses and thermal fluctuations 
enhance cyclization probability of bent polymers.

To summarize, this paper offers an analytical treatment of the mechanics
and thermal fluctuations of twisted and bent polymers. We give a parametric
solution of the problem, which can be used to generate plots of 
experimentally accessible quantities. We expect that this
solvable model would be of  interest to physicists and biologists
interested in the cylclization of DNA.

\begin{acknowledgments} 
\end{acknowledgments}

%


\end{document}